\documentclass[12pt]{iopart}

\begin{document}
\title [Critique on the measurement...]{Critique on the measurement of neutron cross-sections by the Deep
Inelastic Neutron Scattering technique}

\author{ J.J. Blostein, J. Dawidowski\footnote[1]{To whom
    correspondence should be addressed (javier@cab.cnea.gov.ar)} ,
  J.R. Granada }

\address{Consejo Nacional de Investigaciones Cient{\'\i}ficas y T{\'e}cnicas,
  Centro At{\'o}mico Bariloche and Instituto Balseiro, Comisi{\'o}n Nacional
  de E\-ner\-g{\'\i}\-a At{\'o}mica, Universidad Nacional de Cuyo, (R8402AGP)
  Bariloche, Argentina}

\begin{abstract}
  We analyze a recent work of Mayers and Abdul-Redah,[J.Phys.:
  Condens. Matter {\bf 16}, 4811 (2004)] in which the autors report
  the existence of anomalous neutron cross sections in several
  systems. In the present work we show that the Deep Inelastic Neutron
  Scattering (DINS) results presented by the authors are affected by
  an inaccurate formalism employed for obtaining nuclear momentum
  distributions, and therefore definitive conclusions cannot be drawn
  on the subject of anomalous cross sections. We also show the reasons
  why the exact formalism for obtaining momentum distributions that we
  recently published must be employed for analysing the experimental
  data instead of the approximations performed in the mentioned work.
  We also point out serious inconsistencies between different results
  reported in the mentioned work, as well as incompatibilities with
  previous results published by the authors.  These inconsistencies,
  as well as experimental evidence against the existence of anomalous
  cross sections not considered by the authors, reinforce the need to
  revise critically the procedures on which the usual DINS data
  analysis is based as well as the proper characterisation of the
  experimental set-up.
 
\end{abstract}

\maketitle

In a recent work J. Mayers and T. Abdul-Redah \cite{1} presented
experimental results obtained with the Deep Inelastic Neutron
Scattering (DINS) technique as evidence of the existence of anomalous
neutron cross sections in several systems. However, as we have shown
in previous papers, the formalism employed by the authors to analyse
the experimental data is based on ill-founded approximations, which
leads to inaccurate results. This fact created a controversial situation
on the real existence of the mentioned anomalous cross sections, which
is treated in detail in references \cite{01Blo, 03Blo, 03Blo2, 04Blo}.
In this comment we will analyse the validity of the conclusions drawn
in Ref. \cite{1}, where the authors minimize the importance of errors
originated in the approximations assumed in the data analysis. As we
will show in this comment, in \cite{1} the authors have omitted
important results and conclusions presented in the mentioned previous
works, and they have not considered other published works where the
addressed issue is analysed \cite{Ramman,ReplyComm, PRB}. In the
following paragraphs, we will itemise the principal reasons wherefore
the results and conclusions presented in \cite{1} are flawed, as well
as the most important omissions we found in the mentioned work.

\begin{enumerate}

\item Ref. \cite{1} is an important improvement in order to reach a
  general agreement on the way that DINS experimental data must be
  treated. It is worth remembering that in previous works \cite{02Fie,
    96May, 02May,97Cha, 02Cha} the authors have employed the convolution
  formalism without mentioning its approximated nature. In contrast, in
  reference \cite{1} the authors admit that the usually employed DINS
  data treatment (based in a convolution expression) consists in an
  approximation (CA), that leads to inexact results. However, as a
  first attempt to analyse DINS experimental data in the proper way,
  in Ref.  \cite{1} the authors present Eq.  (2.8) as an alternative
  form of the exact formalism, which was previously deduced and
  analysed in \cite{04Blo}, where we have shown that a wide
  distribution of final energies is operative at every time-of-flight
  in a DINS spectrum, instead of a single well-defined one as it is
  assumed in Eq. (2.8).  Such analysis clearly shows that the basic
  hypothesis of a fixed final energy, on which the Eq. (2.8) is based,
  is wrong.

\item In a recently published work \cite{PRB} we introduced the
  exact formalism that must be employed for obtaining momentum
  distributions by DINS to avoid the inaccuracies provoked by the CA.
  One of the most important results in that work is the demonstration
  that the exact integration kernel strongly depends on the time of
  flight. This result shows that the time-of-flight-independent
  resolution function, usually employed as integration kernel in the
  CA framework, is inaccurate. As a consequence, the central
  expression (2.23) of Ref. \cite{1} (also presented by the authors as
  exact) is not correct and should be replaced by an expression where
  the probability $P$ explicitly depends on time-of-flight $t$.

\item It is important to note one of the main results presented in
  Ref. \cite{1}, namely that the peak intensities obtained with the CA
  are strongly dependent on the momentum distributions $J(y)$   employed
  in the fitting process. In Fig. 7 the authors show that CA can
  introduce a systematic reduction in the obtained ratio
$\sigma_{\mathrm{H}}/\sigma_{\mathrm{D}}$. In this process Gaussian functions
for the momentum distributions were assumed. On the other hand, in
Fig. 14, we see quite different results when employing non-Gaussian
distributions. The results appear to be strongly dependent on the
momentum distribution assumed in the fitting process. It is very
important to note that these non-Gaussian $J(y)$ were also obtained by
the authors in the CA framework \cite{02Rei}, and therefore are
affected by the already mentioned approximations.  In consequence, the
results presented in Ref. \cite{1} for
$\sigma_{\mathrm{H}}/\sigma_{\mathrm{D}}$ are also affected by CA and a
definitive conclusion on the supposed anomalies of the Hydrogen and
Deuterium cross sections cannot be drawn.

\item It would be most enlightening if the authors presented the
  results for the Hydrogen, Deuterium, and Oxygen momentum
  distributions they obtained for the different mixtures (which they
  did not in Ref.  \cite{1})), and also if they compared such results
  with those previously published. This information should be readily
  available since it is the primary output of DINS. It is also
  opportune to remember that important dynamical features such as the
  mean kinetic energy obtained by DINS, can be independently checked
  by transmission experiments. As an example, in Ref. \cite{03Blo2},
  we obtained Hydrogen, Deuterium, and Oxygen mean kinetic energies on
  H$_{\mathrm{2}}$O and D$_{\mathrm{2}}$O in excellent agreement with
  the well-known reference values presented in Refs.
  \cite{68Nei,74But}. Furthermore, the results obtained for these
  magnitudes on H$_{\mathrm{2}}$O/D$_{\mathrm{2}}$O mixtures are in
  full agreement with those obtained from linear combination of the
  different molecular species present in such systems
  (H$_{\mathrm{2}}$O, D$_{\mathrm{2}}$O and HDO). Since the momentum
  distributions assumed in the DINS fitting process have a great
  influence on the obtained peak intensities, in the future, more
  efforts should be directed to improve the determination of such
  distributions. For this purpose, the exact formalism recently
  presented in Ref. \cite{PRB} should be the most adequate tool.
  
\item The total cross sections of the employed filters are neither
  Lorentzian nor Gaussian as the authors pointed out. Around each
  resonance, in the simplest approximation, these are given by the
  Lamb equation \cite{64Bec}\footnote{A more detailed description
  should considered a more realistic nuclear momentum distribution in
  the filter, where the gas model assumed in the Lamb equation should
  be replaced by a more detailed model for the solid sate dynamics.}.
  For a more accurate description, the authors could employ the
  neutron cross sections available in \cite{ENDF}, which were
  experimentally obtained in a wide energy range.
  
\item The authors omit to mention that the resolution function
  $R(y)$ they employ is not compatible with the definition of
  resolution function. The resolution function they employ does not
  reproduce the neutron Compton profile in the case of an ideal gas at
  T= 0 K. In \cite{01Blo} we have shown that the only resolution
  function mathematically compatible with the definition of resolution
  can be analytically deduced from the instrument characteristics.
  Anyway, we have shown that this resolution function also leads to
  wrong results, which shows that the problem does not reside in the
  employed resolution function, but in the convolution formalism itself.

\item In Ref. \cite{1} the authors mentioned that the reported
  anomalous cross-sections exhibit different angular behaviours on
  different systems. For example, in
  H$_{\mathrm{2}}$O/D$_{\mathrm{2}}$O mixtures the reported anomaly is
  almost independent of the scattering angle, while in other systems
  it strongly depends on the scattering angle. In this context, it is
  important to note the different and inconsistent ways to process the
  data reported in the literature. More specifically, in the CA
  framework, the factor $v_{\mathrm{1}}/v_{\mathrm{0 }}$ was
  apparently employed in two different ways. On one hand, Refs.
  \cite{02Fie,96May} mention a calculation of this factor based on
  kinematic conditions (\textit{i.e.} $v_{\mathrm{1}}$ is assumed
  fixed by the filter and $v_{\mathrm{0}}$ is calculated for each
  time-of-flight independently of the scattering angle); on the other
  hand Eq. (3.1) of Ref. \cite{1} (see also Eq. (1) of Ref.
  \cite{99Kar}) mentions that this factor was calculated according to
  dynamic conditions (which only depends on the scattered mass and the
  scattering angle, and are independent of the analysed time-of-flight
  and the characteristics of the filter). Such discrepancy in the
  employed data treatments, not mentioned by the authors, could affect
  the different angular behaviours above mentioned.
  
\item In Fig. 10 the results clearly show a gross systematic
  difference between the cross section ratio obtained with the single
  difference method and the one obtained with the double difference
  method (being the first one about 30\% systematically greater than the
  second one). The authors do not explain the origin of these differences.
  If the CA employed by the authors were exact the mentioned
  discrepancies should not exist, since such ratio of neutron cross
  sections is a constant physical magnitude. Such discrepancy casts
  serious doubts not only on the formalism employed by the authors for
  analysing the experimental data, but also on the characterisation
  performed on the experimental set-up. Furthermore, in Fig. 10 the
  authors attempt to show that the overlap effect between different
  peaks is not relevant. However they omit to mention that in
  \cite{03Blo} we have shown that the inaccuracies of CA for obtaining
  peak intensities are still present even when the intensity of an
  isolated peak is analysed, {\it i.e.} when the overlap effect is
  absent.
  
\item Some results presented by the authors are incompatible with
  those published in previous works. For example, in reference
  \cite{02Cha} it is concluded that the reduction reported for the
  ratio $\sigma_{\mathrm{H}}/\sigma_{\mathrm{D}}$ in
  H$_{\mathrm{2}}$O/D$_{\mathrm{2}}$O mixtures is only originated in a
  reduction of $\sigma_{\mathrm{H}}$, while the obtained value of
  $\sigma_{\mathrm{D}}$ agrees with the tabulated value. On the other
  hand, in Fig. 6 opposite results are presented, {\it i.e.} an
  anomaly in the Deuterium intensity and not in Hydrogen. 

\end{enumerate}

The validity of the DINS results presented in \cite{1} not only
depends on the formalism employed for analysing the experimental data,
but also on a proper characterisation of the experimental set-up. In
\cite{1} the authors have performed a characterisation of different
components of the experimental set-up, which in the light of the
inconsistencies observed, could be improved, verified, and/or
performed by alternative methods never employed in VESUVIO.  Details
of such alternative characterisation, as well as additional
disagreements between different results presented in \cite{1}, will be
discussed elsewhere.

We wish to remember that, in other to investigate the cross sections
of Hydrogen and Deuterium in light water/heavy water mixtures, we
performed transmission experiments on such mixtures on the epithermal
neutron energy range employing the Bariloche Electron LINAC
\cite{03Blo2}. Our experiment shows no traces of anomalous neutron
cross sections, and the values we obtained are in perfect agreement
with the tabulated data. As was explained in that work, our
transmission results are conclusive evidence on the absence of
anomalous neutron cross sections in the mentioned mixtures. It must be
noted that both techniques, transmission and DINS, measure the same
magnitude in exactly the same sample and scattering conditions. Due to
the reason exposed in \cite{03Blo2}, any anomaly in the bound-atom
cross section observed by DINS should also be observed in transmission
experiments. The authors of Ref. \cite{1} neither mention that
results, nor the arguments on this subject presented in Ref.
\cite{PRLReply}, where we have shown that the assertions of Ref.
\cite{comm} are wrong.

Finally, it is worth to mention that the absence of anomalous neutron
cross sections in H$_{\mathrm{2}}$O/D$_{\mathrm{2}}$O mixtures was
very recently confirmed by scattering experiments carried out by an
independent group \cite{05Mor}. These results agree with those we obtained
by transmission \cite{03Blo2}, as well as with those obtained by precise
interferometric techniques \cite{99Iof}. In summary, in order to analyse
the anomalies reported/suggested in \cite{97Cha}, three independent
techniques were applied by three independent groups, and
always-negative results were obtained.

\end{document}